\documentclass{emulateapj}
\usepackage{graphicx}
\newcommand{\be}{\begin{equation}}
\newcommand{\ee}{\end{equation}}
\newcommand{\bea}{\begin{eqnarray}}
\newcommand{\eea}{\end{eqnarray}}

\begin{document}

\title{Numerical study of Cosmic Ray Diffusion in MHD turbulence}
\author{A. Beresnyak\altaffilmark{1}, H. Yan\altaffilmark{2}, A. Lazarian\altaffilmark{1}}

\altaffiltext{1}{Astronomy Department, University of Wisconsin, Madison, WI 53706}
\altaffiltext{2}{Astronomy Department, University of Arizona}

\begin{abstract}
  We study diffusion of Cosmic Rays (CRs) in turbulent magnetic
  fields using test particle simulations. Electromagnetic fields
  are produced in direct numerical MHD simulations of turbulence
  and used as an input for particle tracing, particle feedback on
  turbulence being ignored. Statistical transport coefficients
  from the test particle runs are compared with earlier
  analytical predictions. We find qualitative correspondence
  between them in various aspects of CR diffusion. In the
  incompressible case, that we consider in this paper, the
  dominant scattering mechanism occurs to be the non-resonant
  mirror interactions with the slow-mode
  perturbations. Perpendicular transport roughly agrees with
  being produced by magnetic field wandering.

\end{abstract}


\section{Introduction}

The interaction between Cosmic Rays, highly energetic charged
particles, and astrophysical fluids is mediated by magnetic
fields.  As magnetic fields are usually turbulent, CRs do not
freely stream along these fields but scatter (see, e.g.,
Schlickeiser 2002).  Efficient scattering is essential for a
variety of acceleration mechanisms of CRs, such as, for example,
diffusive shock acceleration (Krymsky 1977, Bell 1978, Malkov \&
Drury 2001 and ref. therein).

Understanding MHD turbulence is essential for the correct
description of CR propagation. One popular model has been based
on the combination of slab and two-dimensional perturbations (see
Bieber, Smith, \& Matthaeus 1988). Simplicity of this empirical
model has appealed to researchers and has been used to account
for propagation of CRs in solar wind and magnetosphere.
Numerical simulations (see Cho \& Vishniac 2000, Maron \&
Goldreich 2001, M\"uller \& Biskamp 2000, Cho, Lazarian \&
Vishniac 2002, Cho \& Lazarian 2002, 2003), however, do not show slab modes,
instead, they show Alfv\'enic modes that exhibit scale-dependent
anisotropy consistent with predictions in Goldreich \& Sridhar
(1995, henceforth GS95).  The scalings of compressible modes is
still a subject of debate, although it is suggested that slow
mode is passively advected by Alfven mode (GS95, Lithwick \&
Goldreich 2001), which was verified by numerics, also fast mode showed
relative isotropy which was suggestive of a separate
acoustic-type cascade (Cho \& Lazarian 2002, 2003).

While particular aspects of the GS95 model, e.g. the value of the
spectral index, has been debated (see Boldyrev 2005, 2006,
Beresnyak \& Lazarian 2006, Gogoberidze 2007, Beresnyak \&
Lazarian 2009a,b), this model provide a good start for studying
CR scattering. This program was realized in a number of
publications such as Chandran (2000), Yan \& Lazarian (2002,
2004, henceforth YL02, YL04, respectively), Brunetti \&
Lazarian (2007). In the last three papers, following Cho \&
Lazarian (2003), MHD turbulence has been decomposed into
Alfv\'en, slow and fast modes.

In a complex problem of propagation and acceleration of CRs
we often use so-called diffusive approximation which assume that
the particle scatter or gain energy in small steps. In this
approximation the local particle dynamics will be averaged
to obtain the spatial diffusion coefficient, $D_{xx}$,
and the momentum diffusion  coefficient, $D_{pp}$,
that go into the advection-diffusion equation for the
evolution of quasi-isotropic CR distribution function $f$:

\def\pder#1#2{\frac{\partial #1}{\partial #2}}

\begin{eqnarray}
\pder ft+u\pder fx & = &\pder{}{x}\left(D_{xx}\pder fx\right)\nonumber\\
                   & + & \frac p3 \pder ux \pder fp 
+ \frac 1{p^2} \pder{}{p}\left(p^2 D_{pp} \pder  fp\right),
\end{eqnarray}

(e.g., Skilling 1975). The source terms has be added to the RHS of
this equation to account for injection from thermal particles
and the proper boundary conditions should be defined. Here we
assumed for simplicity that $f(x,p)$ depends only on one
spatial coordinate $x$ and the magnitude of CR momentum, $p$.
This equation uses ``local'' system of reference,
where particle momentum is measured with respect to the rest
frame of the fluid. In a situation when the advection-diffusion
equation is not adequate one has to fall back to more general
approaches, such as Vlasov's equation (see, e.g., Schlickeiser 2002).
In this paper we study particle dynamics assuming
diffusion approximation and we monitor if this dynamics
looks like a diffusive dynamics or not.

The propagation of CRs is a mature quantitative field, which
makes use both of analytical studies and numerical simulations.
For example, a quasi-linear theory (QLT) was used to calculate
scattering of CRs propagating in a mean magnetic field with small
perturbations. However, as turbulence paradigms were changing, so
were the results of CR scattering theories. The purpose of this
paper is to measure CR scattering numerically, based on the best
available direct numerical simulations of MHD turbulence and
compare these results with what scattering theories predict.

QLT has demonstrated that the gyroresonance in GS95 type
turbulence is substantially suppressed and negligible (Chandran
2000, YL02, YL04).  However, the key assumption of QLT, that the
particle's orbit is unperturbed, significantly limits its
applicability. Additionally, QLT has problems in treating
scattering of particles with momentum nearly perpendicular to the
magnetic field (see Jones, Birmingham \& Kaiser 1973, 1978;
V\"olk 1973, 1975; Owens 1974; Goldstein 1976; Felice \& Kulsrud
2001) and perpendicular transport (see K\'ota \& Jokipii 2000,
Matthaeus et al. 2003).

Various non-linear theories have been proposed to improve the QLT
(see Dupree 1966, V\"olk 1973, 1975, Jones, Kaiser \& Birmingham
1973, Goldstein 1976).  In the recent paper of Yan \& Lazarian
(2008, henceforth YL08), a nonlinear formalism (NLT) based on V\"olk
(1975) was developed.  The gyroresonance was found to be marginal
in incompressible turbulence. However, transit time damping
(TTD) was fairly efficient which is different from the QLT
result.  TTD due to nonlinear scattering can be understood as a
scattering by large-scale magnetic compressions (magnetic bottles
formed by slow mode).  The ideas on perpendicular diffusion has been
dominated by the field line random walk (Jokipii 1966, Jokipii \&
Parker 1969, Forman et al. 1974).  It can be justified in a
situation when CRs do not scatter backwards.  However, in
three-dimensional turbulence, parallel transport is also
diffusive, and this can reduce perpendicular transport.

The difference between QLT and NLT has important astrophysical
consequences. Indeed, in some phases, such as hot ISM,
the fast mode is strongly damped, which leaves only Alfven
and slow modes for scattering.
According to the QLT, however, these modes do not provide any significant
scattering. This would predict that there are large volumes in the
disk of the Galaxy where CRs do not scatter at all, which would question
global simulations of propagation of CRs in the Galaxy or halo
made without such an assumption. This will also somewhat contradict
the isotropy of galactic CRs observed on Earth, because isotropy
suggests efficient scattering. Fortunately, NLT corrects this
``zero-scattering'' QLT prediction, putting a lower limit
on the efficiency of CR scattering, thus mitigating contradictions
described above.

%Although NLT prediction puts a lower limit on CR scattering
%in most realistic astrophysical circumstances, there several mechanisms
%that could be more efficient that the ones 
Test particle simulation has been used to study CR scattering and
transport before, see, e.g., Giacalone \& Jokipii (1999), Mace et al (2000), Qin at
al. (2002). The aforementioned studies, however, used synthetic
data for turbulent fields, which have several disadvantages.
Creating synthetic turbulence data which has scale-dependent
anisotropy with respect to the local magnetic field (as observed
in Cho \& Vishniac 2000 and Maron \& Goldreich 2001) is difficult
and has not been realized yet, as far as we know.  Also,
synthetic data normally uses Gaussian statistics and
delta-correlated fields, which is hardly appropriate for
description of strong turbulence. In contrast, in this paper we
are using the results of direct numerical MHD simulations as the
input data for particle scattering simulations.

Another challenging problem is the back-reaction of CRs to the
fluid.  A particular mechanism for such a process, called
streaming instability, has been popular in describing CR
scattering since long time ago (see, e.g., Kulsrud \& Pearce
1969).  In this paper, however, we consider only {\it test
  particle} scattering.  This describes an important physical
limit where CR density is negligibly small and collective effects
are unimportant.  Although in realistic astrophysical environments
CR density is never small and CR pressure is always dynamically
important, the understanding of the {\it test particle limit}
will give us a firm ground for future research into a more
general and more difficult problem of mutual interaction of CRs
and MHD fluid.

In this paper we study the scattering by the incompressible
component of turbulence. If the fast mode is present, however,
it will dominate scattering of low energy CRs, as long as the fast mode 
is not effectively suppressed (YL04). Another very efficient mechanism
of scattering is the instability between CR fluid and MHD fluid (see above).
It is present, e.g., when there are strong CR gradients that lead
to streaming.

In what follows, we discuss numerical methods, including DNS of
MHD turbulence and particle tracing technique in \S~2.  We
discuss theoretical expectations for CR scattering in \S~3.  We
provide numerical measurements of scattering in \S~4 and
measurements of space diffusion in \S~5.  We discuss our results
in \S~6.

\section{Numerical Methods}
In order to trace particle trajectories we were using
electromagnetic fields obtained in direct three-dimensional
simulations of MHD turbulence.  For the purpose of this paper we
were using only incompressible simulations for a variety of
reasons. First, we wanted to test those predictions of the theory
that pertain to incompressible case. Second, the incompressible
simulations were performed with pseudospectral code that has
explicit dissipation and, unlike finite-difference code has no
uncertainties due to numerical dissipation. Also, incompressible
simulations have larger inertial range.

\subsection{DNS of turbulence}
We solved incompressible MHD equations,

\begin{equation}
\partial_t{\bf w^\pm}+\hat S ({\bf w^\mp}\cdot\nabla){\bf w^\pm}=-\nu_n(-\nabla^2)^n{\bf w^\pm}+ {\bf f^\pm},
\end{equation}

written in terms of Elsasser variables which are defined in terms
of velocity $\bf{v}$ and magnetic field in velocity units ${\bf
  b=B}/(4\pi \rho)^{1/2}$ as ${\bf w^+=v+b}$ and ${\bf w^-=v-b}$,
$\hat S$ is a solenoidal projection operator, ${\bf f^\pm}$ is
Elsasser forcing.  These are general equations which can be used
for either turbulence with no mean magnetic field (i.e. when the
average of ${\bf w^+-w^-}$ is zero), or in a presence of such a
mean field. In the latter case perturbations of ${\bf w^+}$ can
be seen as the waves propagating oppositely to the magnetic field
direction. Both Alfven and pseudo-Alfven waves propagate with the
same velocity $v_A=B_0/(4\pi \rho)^{1/2}$.

We used pseudospectral code described in more detail in Beresnyak
\& Lazarian 2009(a,b) (henceforce BL09a,b). The pseudospectral
code solves Eq. 2 as ordinary differential equation in time
for each spacial Fourier harmonic, the ``pseudo'' coming from the fact
that nonlinear term is calculated in real space, and then converted
back to Fourier space. The dissipation and divergence-free condition
for velocity and magnetic field are done with simple algebraic
operations in Fourier space. For time integration we use leapfrog
which is time-reversible and numerical dissipation is absent,
because nonlinear term, calculated in this manner, preserve both energy
and cross-helicity. Therefore the only dissipation come from
the explicit dissipation term. The turbulence was
driven by either independent Elsasser driving or by pure velocity
driving (which formally corresponds to $f^+=f^-$). For the
purpose of this paper we used the results of $768^3$ balanced and
imbalanced turbulent simulations from BL09a. Balanced turbulence
corresponds to the well-studied limit, where the rms values of
$w^+$ and $w^-$ are equal. Physically this corresponds to the
situation when the flow of $w^-$
perturbations, which propagate along the mean magnetic field
direction, balances the opposing flow of $w^+$.  The more general
case of imbalanced turbulence is less studied (see BL08, BL09a
and ref. therein), but more likely to be found in nature. This is
due to the fact, that MHD turbulence is often driven by the
strong localized source of perturbations and near the source we
mostly see waves moving away from the source. Such as a solar
wind turbulence near the Sun, which is strongly
imbalanced\footnote{The measurement of the imbalance in the solar
  wind has been possible with the advent of satellites that
  independently measure the velocity and the magnetic field at
  the same point. Similar measurements for other astrophysical
  sources, such as ISM, are yet to be developed.}.  Naturally, we
are also interested in particle scattering in the imbalanced
turbulence, although few theoretical predictions of scattering
exist in this case, if any.

\begin{figure}
\includegraphics[width=0.95\columnwidth]{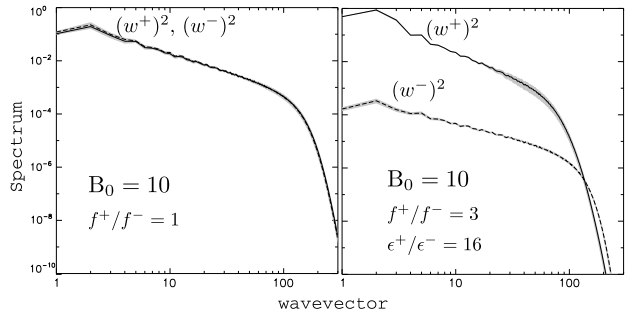}
\caption{Sample spectra from turbulent simulations. Left -- balanced case, right -- imbalanced case.
Shown are Elsasser energy spectra (see BL09a for more details).}
\label{mhdspectra}
\end{figure}

Another dimension in parameter study of BL09a was the strength of
perturbations with respect to the mean field. The $\delta B\sim
B_0$ is called trans-Alfvenic case, where perturbations are of
the order of the mean field. We also consider sub-Alfvenic case
when perturbations were approximately 10 times weaker than mean
field $B_0$, which correspond to so called Alfvenic Mach number
$M_A\sim 0.1$. The latter case can be considered as smaller
scales of trans- or super-Alfvenic turbulence that cannot be
reached directly by 3D simulations of aforementioned flows.  In
order for sub-Alfvenic turbulence to be strong\footnote{ Strong
  MHD turbulence appears naturally as a result of the anisotropic
  cascade. Even if turbulence is driven weakly with respect to
  the mean field, the perpendicular cascade of weak turbulence
  (Galtier et al, 2000) will increase the strength of interaction
  until it becomes strong. The realistic ISM turbulence, however,
  is driven strongly, such as $\delta B\sim B_0$ on the outer
  scale.  So turbulence is strong to begin with and continue to
  be strong along the cascade (GS95).} it has to be driven
anisotropically on its outer scale, which was realized in
BL09a,b. The computational box was also elongated in the
direction of the mean field, with parallel size 10 times larger
than perpendicular size for $M_A=0.1$ case. Throughout the paper,
when we mention the ``box size'' and the ``outer scale of
turbulence'' it means perpendicular size, the parallel size is
the same for trans-Alfvenic cubes and 10 times larger for
sub-Alfvenic cubes.

The note of caution has to be said with regard to sub-Alfvenic
simulations being the small scales of trans-Alfvenic
turbulence. As we use periodic boundaries, the scales which are
larger than the cube size are excluded from consideration. This
means that we cut out a range of scales in the inertial interval
of turbulence and all larger scales are represented only by the
value of the mean magnetic field (the mean velocity can be
excluded by the local frame of reference).  This could or could
not be satisfactory for simulations of particle scattering.  If
the resonant scattering mechanism is effective, then particles
mostly interact with those scales of magnetic perturbations that
are present in the simulation. In the opposite case the
aforementioned interaction can be less effective than the
interaction with large scale perturbations that are not present
in the numerical cubes. In this case the result should not be
trusted. We will return to this question below.  Accidentally,
the QLT consider scattering in a manner which is consistent with
approach that ignores larger scales and consider particle
gyrating along a strong guiding field and interacting with small
resonant perturbations.

The turbulence was driven with specially designed
quasi-stochastic driving on outer scale (with wave numbers in the
interval $k=2..3.5$) with self-correlation time of $2$ in code
units. The driving worked until stationary state was reached. The
scale of the largest coherent eddy in the simulation was around
$0.2$ of the cube size. This is the outer scale of turbulence
$L=0.2$. This largest correlation scale of velocity and magnetic
perturbations is determined by nonlinear interaction and is
typically less than the driving correlation scale. On the driving
scales, i.e. $k=2..3.5$ the turbulence is not yet fully developed
and the spectrum is distorted, having a characteristic bump, and
well-developed turbulence starts with $k=5$. Another definition
of outer scale is through anisotropy.  One can expect the
anisotropy to follow a Goldreich-Sridhar critical balance
$k_\|=k_\perp^{2/3}L^{-1/3}$. This also lead to the estimate of
$L=0.2$[]\footnote{Here we omit $2\pi$ factor normally present in
  size-wavevector relation, as we normalized cube size to
  unity.}, with turbulence being approximately isotropic at $k=5$
and approximately $1:2$ anisotropic at $k=40$ for trans-Alfvenic
case.  At any given time the cube contained large number of
independent turbulent realizations ($>40$). Spectra for one
balanced and one imbalanced case are presented on
Fig.~\ref{mhdspectra}. Further details of the code and simulations
can be found in BL09(a,b).

\subsection{Particle tracing}
The electric field in the laboratory frame was obtained through
$E=-[v\times B]/c$ equation, assuming $v_A/c=10^{-5}$ (a typical
value for the ISM)\footnote{The velocity was measured in the
  Alfvenic units and the electric field is in the same units as
  magnetic field}.  The particles were injected randomly through
the cube and the trajectories were traced by hybrid Runge-Kutta
quality-controlled ODE solver, assuming periodic boundaries for
particles as well as fields.

In particular, we solved 6 equations:

\def\uhat{{\hat {\bf u}}}
\def\shat{{\hat s}}
\def\gamhat{{\hat \gamma}}

\bea
\frac{d\uhat}{\shat}&=&\gamhat{\bf E}+\uhat\times{\bf B}\\
\frac{d{\bf x}}{\shat}&=&r_L\uhat.
\eea

\begin{figure}
\includegraphics[width=\columnwidth]{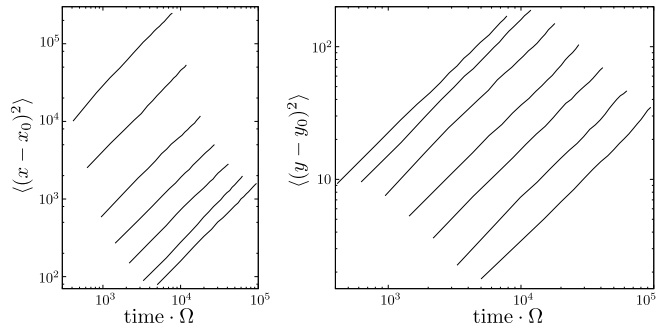}
\caption{Diffusive behavior of the particles in the tracing
  simulations.  For different $r_L$ we show ensemble-averaged
  square deviations, which are proportional to time.  $x$ and $y$
  are measured in units of the cube size.}
\label{xx_yy}
\end{figure}

Here $\uhat$ is the normalized space component of the 4-velocity,
$\uhat={\bf u}/\gamma_0$, where $\gamma_0$ is the initial
particle gamma-factor. Also
$\gamhat=\sqrt{1/\gamma_0^2+\uhat^2}$. $\shat=(eB_0/mc^2)s$ is a
self-time measured in cyclotron frequency units (a gyration
frequency in particle's own frame).  A particle with $\mu=0$ will
make a full orbit in $B_0$ field in $2\pi$ time.  Therefore, we
conveniently measure scattering frequency relative to gyration
frequency.  The measure of initial particle's energy, normalized
Larmor radius is expressed as
$r_L=mc^2\gamma_0/eB_0$. Physically, one can think of $\gamma_0$
is a measure of relativicity of the particle, i.e. for small
$\gamma_0$ we will recover nonrelativistic equations, and for
large $\gamma_0$ -- ultra-relativistic equations. At the same
time, $r_L$ is the measure of energy, but with respect to the
perpendicular size of the simulation box. In most simulations we
took $1/\gamma_0$ (which enters only in the equation for
$\gamhat$) as zero or close to zero, such as $10^{-5}$, this
corresponds to ultra-relativistic particles. The $r_L$ was varied
from 0.1 of the cube size to around a grid size. Fig.~\ref{xx_yy}
presents ensemble-averaged square distance vs time for different
$r_L$. The square distance grows linearly with time, which is
expected for diffusive motion.

\section{Expected CR transport properties in MHD turbulence}

\subsection{Formalism for NLT}

We start with explaining QLT which is the theory for resonant
interactions: gyroresonance scattering and transit scattering
(also called transit time damping, TTD).  The resonant condition
is $\omega-k_{\parallel}v\mu=n\Omega$ ($n=0, \pm 1,2...$), where
$\omega$ is the wave frequency, $\Omega=\Omega_{0}/\gamma$ is the
relativistic gyration frequency, $\mu=\cos\theta$, $\theta$ is
the pitch angle of particles.  TTD corresponds to $n=0$ and it
requires compressible perturbations.  Most of the gyroresonance
contribution comes from $n=1$.

It was demonstrated that scattering by Alfv\'enic turbulence is
substantially suppressed due to its anisotropy (Chandran 2000,
YL02). Fig.~\ref{cartoon} illustrates why interaction is
suppressed.  The scattering rate in GS95 turbulence with outer
scale of $L$ and assuming that $\theta$ is not close to 0 is
given by QLT as (YL02):

\begin{figure}
\hfil
\includegraphics[width=0.6\columnwidth]{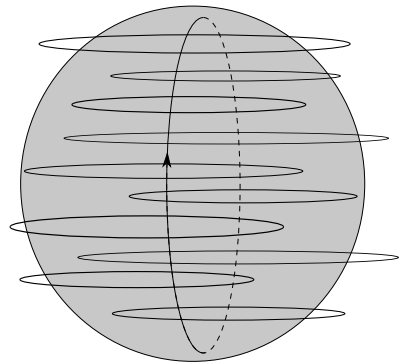}
\hfil
\caption{Cartoon illustrating that the CR gyroresonance
  scattering in strongly anisotropic (GS95) turbulence is very
  ineffective. In perpendicular direction CRs feel many
  uncorrelated eddies and the interaction averages out.}
\label{cartoon}
\end{figure}

\begin{equation}
D_{\mu\mu}=\frac{v^{2.5}\mu^{5.5}}{\Omega^{1.5}L^{2.5}(1-\mu^2)^{0.5}}\Gamma[6.5,k_{max}^{-\frac{2}{3}}k_{\parallel,res}L^{\frac{1}{3}}],\label{ana}
\end{equation}
where $\Gamma[a,z]$ is the incomplete gamma function, $k_{max}$
correspond to the dissipation scale of turbulence and
$k_{\parallel,res}v\mu=\Omega$. The scattering frequency,
therefore, is approximately Bohm-like if Larmor radius is of the
order of $L$, but then falls steeply as $\Omega^{-1.5}$ and
becomes negligible for small energies.

Contrary to QLT which assume that the magnitude of the magnetic
field stay constant, NLT relaxes this assumption and allow this
quantity to change gradually, adiabatically with respect to
particle motion. Due to conservation of adiabatic invariant
$p_\bot^2/B$ (see Landau \& Lifshits 1975) the pitch angle will
gradually vary, resulting in resonance broadening (V\"olk, 1975).
Nonlinear transport (NLT) formalism is based on the replacement
of the sharp resonance between waves and particles
$\delta(k_{\parallel}v_{\parallel}-\omega\pm n\Omega)$ from QLT
to the ``resonance function'' $R_n$ (YL08): \bea
R_n&=&\Re\int_0^\infty dt e^{i(k_\|v_\|+n\Omega-\omega) t-\frac{1}{2}k_\|^2v_\bot^2t^2 \left(\frac{<\delta B_\parallel^2>}{B_0^2}\right)^\frac{1}{2}}\nonumber\\
&=&\frac{\sqrt{\pi}}{|k_\|\Delta v_\||}\exp\left[-\frac{(k_\|v
    \mu-\omega+n\Omega)^2}{k_\|^2\Delta v_\|^2}\right],
\label{resfunc}
\eea

\begin{figure}
\includegraphics[width=0.95\columnwidth]{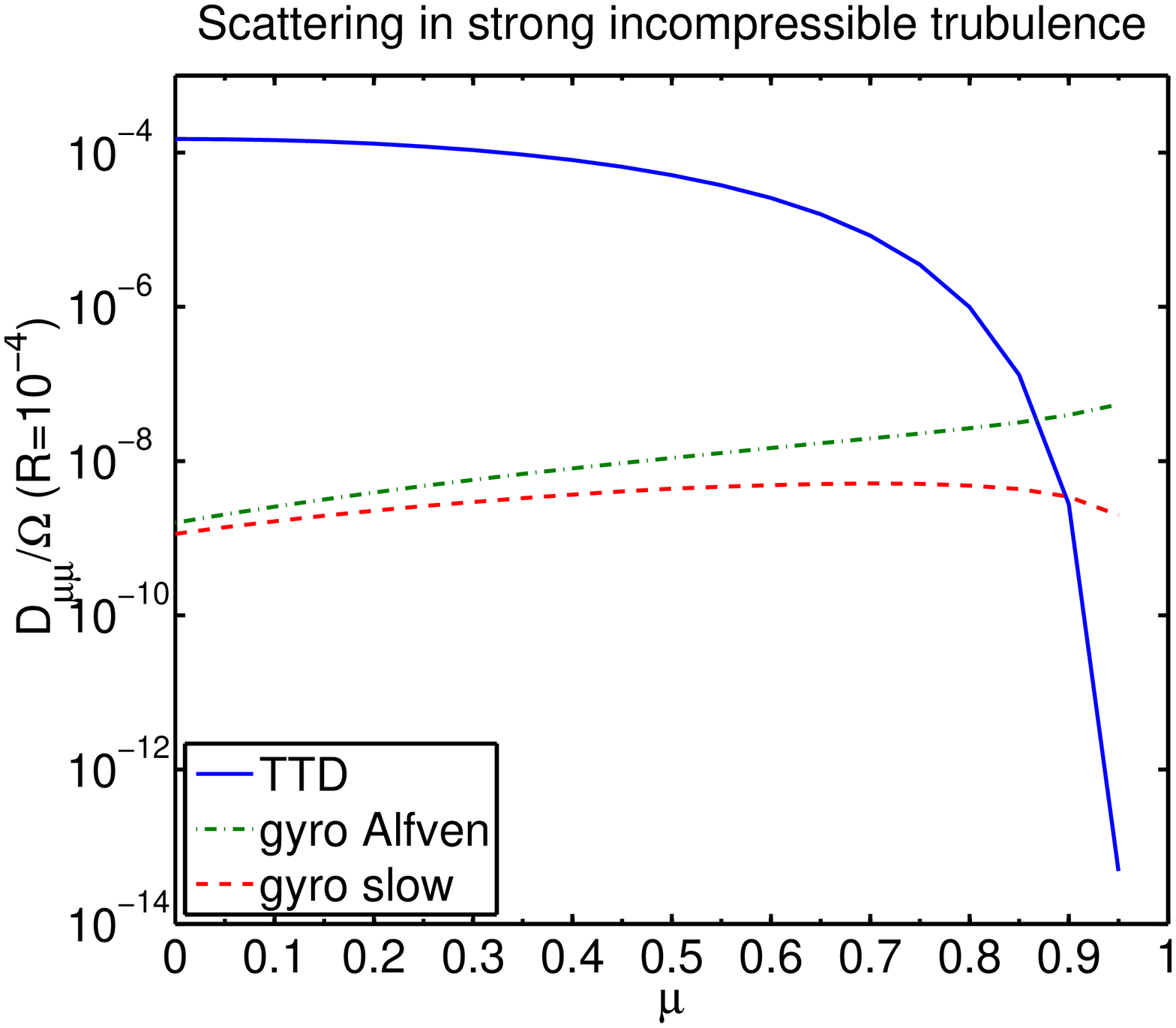}
\caption{Scattering of CRs with $R=r_L/L=10^{-4}$ in strong
  incompressible trans-Alfvenic turbulence. Solid line represents
  TTD contribution. Dash-dot refers to the gyroresonance with the
  Alfv\'en mode while dash line is for the gyroresonance with the
  slow mode. See eqs \ref{TTDstrong}, also see YL08.}
\label{strongturb}
\end{figure}

The width of the resonance function depends on the perturbation
strength of the turbulence $\Delta \mu=\Delta v_\|/v_\bot\simeq
\sqrt{\delta B/B}=\sqrt{M_A}$). For gyroresonance ($n=\pm
1,2,...$) the result depends on whether $\mu$ is strongly or
weakly perturbed by regular field. If $\mu\gg \Delta \mu$, the
result is similar to QLT, because the exponents in
Eq.(\ref{resfunc}) become close to $\delta$-functions.  For
$\mu<\Delta \mu$, however, the result is different. To
demonstrate this we can consider the case of $90^\circ$
scattering. Indeed, if $\mu\rightarrow 0$, the resonance happens
mostly at $k_{\|,res}\sim \Omega/\Delta v$, while in QLT
$k_{\|,res}\sim\Omega/v_\|\rightarrow \infty$. Compared to TTD,
however, $D_{\mu\mu}$ for gyroresonance is still smaller in
incompressible case (see Fig.~\ref{strongturb}) due to
anisotropy.

\subsection{Scattering in strong MHD turbulence}

Assuming the tensor of magnetic perturbations introduced in Cho
et al. (2002), which is consistent with the GS95 model, we can
calculate scattering from TTD and gyroresonance. Assuming small
energies corresponding to Larmor radii much smaller than the
outer scale one gets for TTD (YL08):

\bea
D^T_{\mu\mu}&=&\frac{\sqrt{\pi} M_A^{7/2}v}{16L}(1-\mu^2)^{3/2}\left[-E_1(q\xi_\|)-e^{-q\xi_\|}\right]_{1}^{\xi_{\|,max}}\nonumber\\
&&\times\exp\left[-\frac{(\mu-v_A/v)^2}{\Delta \mu^2}\right],
\label{TTDstrong}
\eea
where $\xi=kL$, $E_1(\xi)=\int_1^\infty dt \exp(-\xi t)/t$
and $q=(\xi_{\bot,max}M_A^2)^{-2/3}$.

Fig.~\ref{strongturb} displays the pitch angle diffusion
coefficients resulting from TTD scattering and gyroresonance
(YL08). We see that gyroresonance is mostly subdominant, however
at small pitch angles TTD is inefficient and gyroresonance
dominates.

\begin{figure}
\includegraphics[width=0.95\columnwidth]{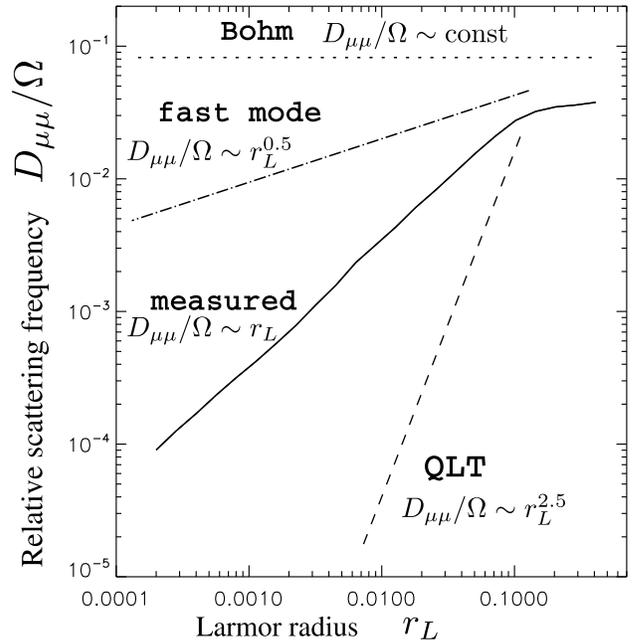}
\caption{Dimensionless CR scattering coefficient $D_{\mu
    \mu}/\Omega$ for the case of $\mu=0.71$ vs Larmor radius
  $r_L$ expressed in cube size units (solid line). We suppose
  that it is dominated by magnetic bottles formed by slow mode,
  this is why $D_{\mu \mu}$ (dimensional scattering frequency) is
  almost constant, i.e. independent on particle's energy. For
  comparison, we plot various theoretical predictions: QLT
  prediction for Alfven and slow mode (dashed); QLT prediction
  for fast mode (dot-dashed), note that in our data fast
  mode was absent, so this line is only for reference;
  hypothetical Bohm scattering or maximally efficient scattering
  (dotted).}
\label{dmumu}
\end{figure}

\section{Measurements of scattering}

$D_{\mu \mu}$ scattering property was measured in the tracing
experiments where an ensemble of particles with the same $r_L$
(energy) and a particular $\mu_0$ were traced by a certain
time. This time was determined by the condition that the rms of
deviations of $\mu$ is small (i.e. 0.1-0.01).  Then the curves of
the ensemble-averaged $\langle(\mu-\mu_0)^2\rangle$ were fitted
with a linear curve, and so $D_{\mu \mu}$ was obtained.  The
$D_{\mu \mu}$ for trans-Alfvenic case is presented on
Fig.~\ref{dmumu}.  For sub-Alfvenic case we noticed that there
were very few $90^\circ$ scattering events. This will be
explained in the next section.

As we see from Fig.~\ref{dmumu} the measurement of scattering
frequency is {\it incompatible} with QLT. The scattering
frequency normalized to the gyration frequency is proportional to
the Larmor radius i.e.  it is constant with energy (as $\Omega
r_L=v \approx c$). It would be reasonable to assume then that
particles of all energies scatter on the same objects, magnetic
bottles, formed by large scale slow-mode perturbations. The same
result could be obtained from NLT, taking into account $\Delta
\mu \sim \mu$ in strong turbulence. At larger energies scattering
becomes less efficient i.e. high energy particles ``feel'' less
mirrors. This transition happen at around $r_L/L\approx 0.1$.

Due to the lack of outer scale, the $D_{\mu \mu}$ for
sub-Alfvenic case is supposed to be QLT-like and very small. As
such it was severely contaminated by numerical error, in
particular, fields interpolation error and was not obtained in
this study.

\section{Measurements of space diffusion}

The measurements of space diffusion $D_{xx}$ and $D_{yy}$
were more straightforward than the measurements of $D_{\mu \mu}$
because we did not limit the integration time as in the previous
section. Therefore, we integrated for as long as it took for
the $\langle(x-x_0)^2\rangle$ and $\langle(y-y_0)^2\rangle$
to show good diffusive linear dependence with time.
Those integration times turned out to be very long, so the
particles crossed outer-scale of turbulence for many times.
Therefore these measurements correspond to diffusion on outer
scale and not to ``sub-diffusion'' (see, e.g., YL08).
Moreover, we measured diffusion with respect to some global
frame of reference, determined by the global mean magnetic field.
Therefore, our measurements do not necessarily correspond
to theories that measure ``parallel'' or ``perpendicular'' diffusion
with respect to the magnetic field lines.

\begin{figure}
\includegraphics[width=0.95\columnwidth]{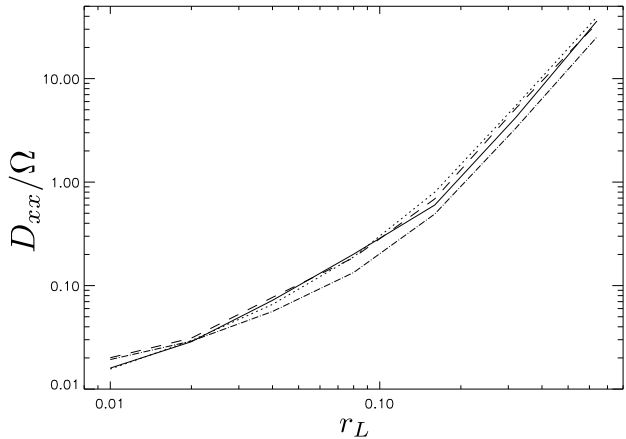}
\caption{Parallel diffusion in trans-Alfvenic case. Solid --
  "balanced", dotted -- "slightly imbalanced" dashed -- "strongly
  imbalanced", dot-dashed -- "very strongly imbalanced" (see
  BL09a for more details).}
\label{Dxx_tra}
\end{figure}

Also, we were only able to obtain {\it the lower limit} of
$D_{xx}$ for {\it sub-Alfvenic} case due to the very low
$90^\circ$ scattering frequency. This was manifested by the fact
that as we increased the precision of the code, the $D_{xx}$
increased and did not show convergence. This very low scattering
frequency has to do with what we discussed earlier -- the lack of
larger scale perturbations in sub-Alfvenic cubes. In this case,
since $\Delta \mu< \mu$, the resonance function becomes narrow so
that marginal interaction is available at $90^\circ$\footnote{The
  case with sub-Alfv\'enic turbulence may be viewed as a
  transition to the QLT case, where the resonance function
  shrinks to $\delta$ function and there is no TTD resonance at
  $90^\circ$}. In other words, in the absence of the large-scale
perturbations, which are normally present in nature, but absent
in our sub-Alfvenic cubes, the $90^\circ$ scattering becomes
problematic and parallel diffusion is replaced by ballistic
propagation along mean field.  At the same time, this suggests
that QLT (resonant scattering) can not be used for low energy
particle scattering, as large scales contribute more than the
resonant scales.

As the mean magnetic field was along 'x' axis, our $D_{xx}$
coefficient correspond to ``parallel diffusion'', while $D_{yy}$
correspond to ``perpendicular diffusion''. This correspondence,
however, is tentative, since most theories predict ``parallel''
or ``perpendicular'' diffusion as happening with respect to the
local magnetic field lines. We nevertheless will use terms
``parallel'' and ``perpendicular'' to $D_{xx}$ and $D_{yy}$. We
also claim that the measurement of the diffusion with respect to
the global reference frame has more practical importance and is
easier applicable to the results of observations.

\subsection{Parallel diffusion}
The results for parallel diffusion for trans-Alfvenic case are
presented on Fig. ~\ref{Dxx_tra}. Along with standard 'balanced'
MHD turbulence case (presented by solid line) we calculated this
diffusion coefficient for simulations with different degree of
imbalance, using datacubes from simulations of Beresnyak \&
Lazarian 2009a.  As the aforementioned paper (along with the
earlier study Beresnyak \& Lazarian 2008) are the first
high-resolution simulations of stationary strong imbalanced
turbulence, it is important\footnote{For example, Solar Wind
  exhibit imbalanced turbulence up to distances of around 1 AU,
  with perturbations coming from the Sun being prevalent over
  backward-going perturbations.} to numerically study the
scattering coefficient, even more so when the theory is lacking.

As we see from Fig.~\ref{Dxx_tra} at small energies
$D_{xx}/\Omega$ is linearly proportional to $r_L$, i.e. as in the
case with $D_{\mu\mu}$ the scattering frequency is independent of
energy. At higher energies it becomes proportional to the square
of $r_L$. This is again consistent with the behavior of
$D_{\mu\mu}/\Omega$ from Fig. ~\ref{dmumu}, i.e.  that at high
energies $D_{\mu\mu}\sim \Omega$ and as $D_{xx}\sim
1/D_{\mu\mu}$, $D_{xx}/\Omega \sim 1/\Omega^2 \sim r_L^2$. The
transition happens at $r_L/L\approx 0.1$, same as in
Fig.~\ref{dmumu}. So we conclude that the measurements of
$D_{xx}$ and $D_{\mu\mu}$ are consistent with each other. A note
of caution towards direct comparison of these two measurements is
due, however. In the measurement of $D_{xx}$ we did not control
particle's energy, which could undergo changes during the long
integration times of $D_{xx}$ measurement. $D_{\mu\mu}$, however,
was measured during short times, and as electric field was
assumed small (smaller than $B$ by a factor of $v_A/c\approx
10^{-5}$), there wasn't significant energy change during this
short time.  This can explain why the transition between two
regimes of scattering is more sharp of Fig.~\ref{dmumu} rather
than Fig.~\ref{Dxx_tra}. Also, as we mentioned previously,
$D_{xx}$ is the diffusion coefficient measured in the {\it global
  reference frame}, while $D_{\mu\mu}$ defines pitch-angle
scattering with respect to the local field direction.

Fig.~\ref{Dxx_tra} also shows that the diffusion coefficient is
pretty independent on the degree of imbalance, indicating that
the trans-Alfvenic imbalanced turbulence has approximately as
many magnetic bottles as its balanced counterpart. This is
consistent with the assumption that only large-scale
perturbations significantly contribute to scattering. Indeed, in
the imbalanced simulations of BL09a the outer-scale magnetic
field was determined primarily by the stronger Elsasser component
and has similar structure and magnitude and outer-scale magnetic
field in balanced simulations.

\begin{figure}
\includegraphics[width=0.95\columnwidth]{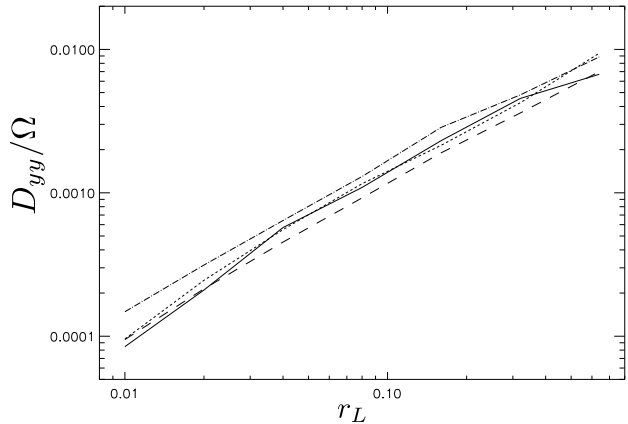}
\caption{Perpendicular diffusion in sub-Alfvenic
case $\delta B/B=1/10$}
\label{Dyy_suba}
\end{figure}

\begin{figure}
\includegraphics[width=0.95\columnwidth]{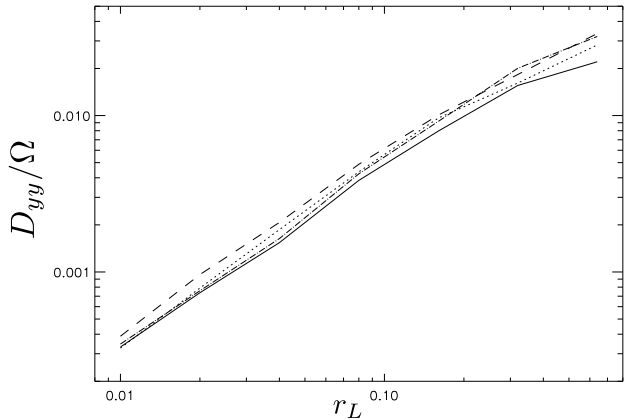}
\caption{Perpendicular diffusion in trans-Alfvenic
case $\delta B/B=1$.}
\label{Dyy_tra}
\end{figure}

\subsection{Perpendicular diffusion}
Perpendicular diffusion coefficients are presented on
Figs.~\ref{Dyy_suba},~\ref{Dxx_tra}.  As to various models,
regimes and terminology of perpendicular diffusion we refer the
reader to YL08 and refs therein.  Let us first interpret the
measurements in the sub-Alfvenic case.  We chose initial
particle's pitch angle to be $45^\circ$.  As we discussed
earlier, due to the particular choice of the data, there wasn't
any $90^\circ$ scattering in this case.  I.e. particles moved
ballistically along 'x' axis, but their trajectories diffused
from the center due to magnetic field wandering. As suggested by
Fig.~\ref{Dyy_suba} the dependence of this plot is almost linear,
i.e. $D_{yy}/\Omega\sim r/L$ or $D_{yy}$ is independent of
energy.

In the three-dimensional turbulence, field lines are diverging
away due to shearing by Alfv\'en modes (see Lazarian \& Vishniac
1999, Narayan \& Medvedev 2002). Most recently the diffusion in
magnetic fields was considered for thermal particles in Lazarian
(2006, 2007).  The cross-field transport can result from the
deviations of field lines at small scales, as well as field line
random walk at large scale.

If we assume that the particle follow magnetic field line and is
diffused only by the outer-scale magnetic field wandering, the
perpendicular diffusion can be expressed as
$D_{yy}/(L^2\Omega)\approx
2^{-1/2}(L_{tr}/L)^2\cdot(L/L_\|)\cdot(r_L/L)$, where $L_\|$ is
the outer parallel scale (which is 10 times bigger than $L$ in
our sub-Alfvenic simulation), $1/\sqrt{2}$ is the cosine of pitch
angle and $L_{tr}$ is the distance the particle is deflected when
it travels $L_\|$ along the field line (see eq. (26) in YL08). We
would expect $L_{tr}$ to be close to $L$.  From the fit of
Fig.~\ref{Dyy_suba} we derive $L_{tr}/L \approx 0.92$ which is
fairly close, considering the uncertainty in $L$.  Using the same
argumentation we obtain $L_{tr}/L \approx 0.53$ from the fit of
Fig.~\ref{Dxx_tra} which is short of what we expected.  This is
an indication that the impediment of travel in parallel direction
which is present in trans-Alfvenic case due to $90^\circ$
scattering decreases diffusion in perpendicular direction. We
stop with this conclusion, as there is clearly not enough data
for a detailed comparison with different models in YL08.

\section{Discussion}
In this paper we numerically measured diffusion coefficients that
arise when particle propagates in a turbulent magnetic fields.
Unlike previous studies, we used realistic fields obtained in a
three-dimensional simulations of MHD turbulence. The focus of
this paper was the incompressible case, where fast magnetosonic
mode is absent. The earlier QLT calculations presumed that
particle scattering is negligible in this case, as the
perturbations are extremely anisotropic with respect to the mean
field. We figured that QLT is not applicable when the magnitude
of the magnetic field is strongly perturbed, and that another
approach called NLT has to be adopted (YL08). NLT allows relatively
efficient scattering through TTD as the particle's pitch angle
changes adiabatically and makes possible for $90^\circ$
scattering. One can interpret this as scattering through
large-scale magnetic mirrors.

We confirmed this picture of mirrors by measuring scattering
frequency which is independent on energy, for small energies. We
also studied spacial diffusion which, in the case of parallel
diffusion is related to scattering frequency. The case of
perpendicular diffusion is more complicated.  We showed that if
particles do not scatter in parallel direction, the perpendicular
diffusion is mostly due to magnetic field line wandering.  This
case could be unphysical though, as the absence of parallel
scattering was due to the absence of larger scales in the
simulation. In the case when parallel diffusion was operating,
the perpendicular diffusion was reduced. At this point, however,
we don't have enough data to distinguish between different models
of perpendicular diffusion.

The special attention should be brought to the astrophysical interpretation
of scattering in the imbalanced turbulence. Figs. 6-8 indicate that
the scattering is similar to the balanced case. This is qualitatively
and quantitatively agree with the picture that was presented in this paper,
namely that in the incompressible case most of the scattering
will come from the outer scale of turbulence and most of the perpendicular
diffusion will come from the field wandering on the outer scale.
This fact, however, does not mean that the scattering in the astrophysical
objects will be the same regardless of the degree of imbalance.
The key to understand this is to understand the nature of our MHD simulations.
In these simulations we kept the fluctuation amplitude and the anisotropy
{\it controlled on outer scale}, the physically all-important {\it dissipation
rate}, however, varied greatly depending on the degree of imbalance.
In astrophysics, turbulence is caused by the sources of kinetic energy,
such as stellar and AGN jets, stellar winds interacting with the ISM,
supernovae, 
the Sun creating perturbations in the solar wind. Turbulent dissipation
will have to balance this influx of energy, however, dissipation
depends greatly on the degree of imbalance, therefore, in a situation
with a constant influx of energy imbalanced turbulence will have much
larger perturbation amplitudes, which will result in a much more efficient
scattering. With respect to relation between dissipation rate and perturbation
amplitude we refer to the imbalanced turbulence model presented
in Beresnyak \& Lazarian 2008, as the most realistic model to-date,
and the simulations in BL09a. A word of caution towards directly
using these results is due, however. Aforementioned model and the 
simulations in BL09a describe {\it stationary} imbalanced turbulence.
However, as we learned from these studies, the time of establishment
of stationary state greatly increases with larger imbalances.
As astrophysical processes are usually transient, it is possible
that in a situation with large imbalance the stationary state will
not be achieved. The {\it stationary} imbalanced turbulence
could still be used to infer the propertied of small-scale fluctuations,
as timescales are smaller, but, as we saw in this study, large scales
are important for scattering. This problem will be solved by the models
of transient and inhomogeneous imbalanced turbulence, although at present
such models are still in their infancy (see, e.g., the Appendix in BL09a).
We are optimistic, however, that the properties of CR scattering in
realistic astrophysical objects that feature imbalanced turbulence,
such as solar and stellar winds, AGN jets and many others
will be figured out.

Although NLT prediction for nonresonant mirror scattering by incompressible turbulence
puts a lower limit on CR scattering, in most realistic astrophysical circumstances,
there several mechanisms that could compete with it.
If the fast mode is present and have a sufficient amplitude in the range
of scales corresponding to Larmor radii of low-energy CRs,
those CRs will be scattered primarily
due to fast mode (YL04). Also, if CRs have large density gradients and tend to stream
in a particular direction, such as CRs escaping the Galaxy, or CRs streaming in front
of the supernovae shock, the backreaction to MHD fluid will be important.
Speaking of turbulence, in this paper we considered generic astrophysical turbulence driven on
large scales. Other types of astrophysical turbulence are often important for scattering
and acceleration. This includes MRI turbulence (see, e.g., Hawley et al, 2001),
and turbulence generated by CR-MHD fluid interaction in supernova shocks
(see, e.g., Beresnyak et al 2009 and ref. therein).
%\citep[see, e.g.,][and ref. therein]{BJL09}.

Stochastic acceleration by MHD turbulence was not studied here as
the correct calculation requires {\it time-dependent} MHD fields,
so that simulations of turbulence are integrated at the same time
as particles propagate. This will be a matter of a future research.

\begin{acknowledgments}
  AB is supported by ICECUBE project. AB is grateful to TeraGrid project
  for providing him with computational resources. HY is supported by Arizona
  Prize Fellowship. AL acknowledges the NASA grant X5166204101,
  the NSF grant ATM-0648699, as well as the NSF Center for
  Magnetic Self-Organization in Laboratory and Astrophysical
  Plasmas.
\end{acknowledgments}

\end{document}